\documentclass[pdflatex,sn-mathphys-num]{sn-jnl}

\usepackage{graphicx}
\usepackage{multirow}
\usepackage{amsmath,amssymb,amsfonts}
\usepackage{amsthm}
\usepackage{mathrsfs}
\usepackage[title]{appendix}
\usepackage{xcolor}
\usepackage{textcomp}
\usepackage{manyfoot}
\usepackage{booktabs}
\usepackage{bookmark}
\usepackage{algorithm}
\usepackage{algorithmicx}
\usepackage{algpseudocode}
\usepackage{listings}
\usepackage{physics}
\usepackage{siunitx}
\usepackage{bbold}
\usepackage[T1]{fontenc}
\let\orcidlogo\relax
\usepackage{orcidlink}

\theoremstyle{thmstyleone}

\theoremstyle{thmstyletwo}

\theoremstyle{thmstylethree}

\graphicspath{{Figures/}}

\AtBeginDocument{\RenewCommandCopy\qty\SI}
\NewCommandCopy\siunitxqty\qty

\begin{document}

\title[Quantum Volunteer's Dilemma on NISQ Hardware]{Experimental Implementation of the Quantum Volunteer's Dilemma on NISQ Hardware: Noise Analysis and Digital-Twin Validation}

\author*[1]{{\fnm{Germ\'{a}n~D.} \sur{D\'{i}az~Agreda}}\,\orcidlink{0009-0004-0032-0618}}
\email{germandiaz@unicauca.edu.co}

\author[1]{{\fnm{Jhon~Alejandro} \sur{Andrade~Hoyos}}\,\orcidlink{0000-0003-1406-3064}}
\email{jaandrade@unicauca.edu.co}

\author[1]{{\fnm{Carlos Andr\'{e}s} \sur{Dur\'{a}n}}\,\orcidlink{0009-0008-3243-7684}}
\email{caduran@unicauca.edu.co}

\author[2]{{\fnm{Sebasti\'{a}n Andr\'{e}s} \sur{Cajas~Ordo\~{n}ez}}\,\orcidlink{0000-0003-0579-6178}}
\email{sebasmos@mit.edu}

\author[3]{\fnm{Noah Dane} \sur{Hebdon}}
\email{nhebdon2@jh.edu}

\author[4,5]{{\fnm{Siong~Thye} \sur{Goh}}\,\orcidlink{0000-0001-7563-0961}}
\email{gohst2@a-star.edu.sg}

\author[3,4,6]{{\fnm{Dax~Enshan} \sur{Koh}}\,\orcidlink{0000-0002-8968-591X}}
\email{dax\_koh@a-star.edu.sg}

\affil[1]{\small \orgdiv{Department of Physics}, \orgname{Universidad del Cauca}, \orgaddress{\city{Popay\'{a}n}, \country{Colombia}}}

\affil[2]{\small \orgdiv{MIT Critical Data}, \orgname{Massachusetts Institute of Technology}, \orgaddress{\city{Cambridge}, \state{MA}, \country{USA}}}

\affil[3]{\small \orgdiv{Quantum Innovation Centre (Q.InC)}, \orgname{Agency for Science, Technology and Research (A*STAR)}, \orgaddress{2 Fusionopolis Way, Innovis \#08-03, \city{Singapore} 138634, \country{Republic of Singapore}}}

\affil[4]{\small \orgdiv{Institute of High Performance Computing (IHPC)}, \orgname{Agency for Science, Technology and Research (A*STAR)}, \orgaddress{1 Fusionopolis Way, \#16-16 Connexis, \city{Singapore} 138632, \country{Republic of Singapore}}}

\affil[5]{\small \orgdiv{Lee Kong Chian School of Business}, \orgname{Singapore Management University}, \orgaddress{50 Stamford Rd, \city{Singapore} 178899, \country{Republic of Singapore}}}

\affil[6]{\small \orgdiv{Science, Mathematics and Technology Cluster}, \orgname{Singapore University of Technology and Design (SUTD)}, \orgaddress{8 Somapah Road, \city{Singapore} 487372, \country{Republic of Singapore}}}

\abstract{\unboldmath We present an experimental implementation of the multiplayer Quantum Volunteer's Dilemma on noisy intermediate-scale quantum (NISQ) hardware, executed on the \texttt{ibm\_kingston} backend via Qiskit Runtime. The game is evaluated for $N = 2$ to $9$ players under four transpiler optimization levels, with 20 independent repetitions per configuration and 2048 shots per circuit, including post-processing readout error correction via \texttt{mthree}.

Target-state fidelity decays with system size but remains above 70\% (corrected) through $N = 9$. With readout correction, the global average payoff reproduces the quantum theoretical benchmark exactly for $N \leq 6$ and exceeds the classical Nash equilibrium across the full tested range. Optimization level 2 is selected as the reference configuration after gate count analysis reveals that levels 2 and 3 produce identical transpiled circuits, with level 2 achieving superior fidelity stability.

A Hamming distance analysis of raw measurement counts shows that single-qubit errors dominate at small $N$, with multi-qubit contributions growing beyond $N = 6$. A calibration-based digital twin captures global payoff trends but exhibits a linear fidelity decay profile that diverges from the hardware behavior at large $N$, exposing the limits of first-order independent per-qubit noise models.

These results demonstrate that aggregate quantum advantage in multiplayer games is robust to NISQ noise conditions across the full tested range, while the practical observability of state-level advantage is constrained to $N \leq 8$ under post-processed readout correction.}


\maketitle

\section{Introduction}\label{sec:introduction}

    The Volunteer's Dilemma~\cite{bib6} is a classic problem in game theory in which each individual in a group of $N$ players must simultaneously decide whether to ``volunteer'', incurring a personal cost to ensure a collective benefit, or ``abstain'', free-riding on others' contributions. It suffices for at least one player to volunteer for the group to obtain the benefit, which creates a fundamental strategic tension: each agent prefers that someone else bear the cost \cite{diekmann1986volunteers,mercade2021volunteers,otsubo2008dynamic}. From a classical perspective, this dilemma leads to inefficient equilibria, particularly as the number of players grows \cite{goeree2017experimental}. 

Quantum game theory alters this dynamic by expanding the space of allowed strategies \cite{bib4,bib3}. By introducing quantum resources such as superposition and entanglement, quantum games allow players to employ non-classical strategies that can lead to equilibrium structures unattainable by their classical counterparts~\cite{khan2018quantum,eisert2000quantum}. Since the pioneering works of Meyer \cite{bib4} and Eisert--Wilkens--Lewenstein (EWL) \cite{bib3}, this paradigm has been extended to a broad class of game-theoretic scenarios~\cite{du2000nash,du2002playing,piotrowski2002quantum,maioli2018quantization,szopa2021efficiency,kastampolidou2023quantum,frkackiewicz2024permissible,andronikos2025ghz,bib1,khan2025quantum,frackiewicz2025permissible_four,tiago2025classical,weeks2025quantum,allah2026possibility,ahmad2026native,essalmi2026multi-player}, generalized settings \cite{benjamin2001multi-player,du2002mutli,li2002continuous,ikeda2021infinitely}, and studies of how quantum resources \cite{du2001entanglement,du2002entanglement,li2014entanglement,wei2017quantum,mohamed2023quantum} and noise \cite{johnson2001playing,chen2003quantum,flitney2004quantum,shuai2007effect,huang2016quantum,khan2018dynamics,kairon2020noisy,legon2023joint} affect quantum strategic advantage. Experimental implementations on various platforms have further demonstrated quantum games as useful testbeds for near-term quantum devices \cite{lu2004linear,buluta2006quantum,mitra2007experimental,schmid2010experimental,xu2022experimental,agreda2025bridging}.

Within this broader line of work, Koh, Kumar and Goh~\cite{bib1} recently introduced a quantum formulation of the dilemma, the Quantum Volunteer's Dilemma, within the EWL framework~\cite{bib3}. Their analysis shows that quantum strategies can reach Nash equilibria with higher expected payoffs than those attainable through classical mixed strategies, in particular under the cost-sharing variant studied by Weesie and Franzen~\cite{bib7}. However, these results are derived under idealized assumptions. A natural question then arises: does this quantum advantage persist when the game is implemented on real quantum hardware?

The present work is informed by our prior experimental implementation of the Battle of the Sexes on IBM Quantum hardware~\cite{agreda2025bridging}, which demonstrated that EWL-based quantum games are practically viable on NISQ devices and that hardware noise does not necessarily eliminate quantum advantage. Here we extend this line of empirical work to the multi-player Volunteer's Dilemma for $N = 2, \ldots, 9$, introducing a systematic transpiler optimization analysis and a calibration-based digital twin that were not part of that earlier study.

Current quantum devices operate in the noisy intermediate-scale quantum (NISQ) regime~\cite{bib5}, characterized by significant noise including gate errors, decoherence, readout errors, and unwanted couplings between qubits \cite{cheng2023noisy}. These effects can degrade the quantum correlations responsible for the theoretical advantage, so their impact must be explicitly assessed.

In this work we address this question through an experimental approach complemented by data-driven noise modeling. We implement the Quantum Volunteer's Dilemma for $N = 2, \ldots, 9$ on IBM's \texttt{ibm\_kingston} quantum processor, systematically exploring four transpiler optimization levels and analyzing the fidelity of the target state, noise redistribution, and payoff behavior. We additionally construct a \textit{digital twin} of the device based on real calibration data, which we use to compare the model predictions against experimental results quantitatively.

Our results show that, although the fidelity of the target state decreases with system size, the global average payoff holds up well: it remains above the classical Nash equilibrium for all $N = 2, \ldots, 9$, with margins of approximately $0.16$--$0.17$ at $N = 2$ and $0.24$--$0.28$ at $N = 9$ (raw and readout-corrected results at optimization level L2).. The quantum advantage at the aggregate payoff level persists even when the state preparation is imperfect. By comparing against the classical equilibrium, we identify the regimes in which this advantage holds and those in which it disappears, and establish practical limits for its observability in multi-agent systems.

\section{Theoretical Framework}\label{sec:theory}

\subsection{The Classical Volunteer's Dilemma}\label{subsec:classical}

The Volunteer's Dilemma describes a situation in which a group of $N$ agents must each decide whether to contribute to a public good. If at least one player volunteers, the collective benefit is obtained; otherwise, all players receive an unfavorable outcome. In the variant considered in this work, the provision cost is not borne by a single individual but is shared among all who decide to volunteer. This structure introduces a balance between individual incentives and collective cooperation, in which the marginal cost of contributing decreases as more players participate.

We consider the classic Volunteer's Dilemma under mixed strategies, following the formulation of~\cite{bib1}. Each player $i$ adopts a probabilistic strategy $\pi_i \in [0,1]$, where $\pi_i$ denotes the probability of volunteering, while $1 - \pi_i$ corresponds to abstaining.

The game can be formalized as an $n$-tuple
\[
G^{(n)}_{\mathrm{MVD}} = (T_1, T_2, \ldots, T_n; \, \$_1, \$_2, \ldots, \$_n),
\]
where $T_i = [0,1]$ is the set of strategies of the player $i$ and $\$_i : [0,1]^n \to \mathbb{R}$ is their expected payoff function, defined as the expected value of the underlying deterministic game:
\[
\$_i^{\mathrm{MVD}}(\pi_1, \ldots, \pi_n)
=
\sum_{x \in \{0,1\}^n}
\$_i^{\mathrm{VD}}(x)\,
q_{\pi_1,\ldots,\pi_n}(x),
\]
where $\$_i^{\mathrm{VD}}(x)$ is the payoff in the deterministic game. Let $w(x) = \sum_{j=1}^n x_j$ denote the number of volunteering players. The deterministic payoff is
\[
\$_i^{\mathrm{VD}}(x) =
\begin{cases}
0 & w(x) = 0, \\
b & x_i = 0,\; w(x) > 0, \\
b - c/w(x) & x_i = 1,\; w(x) > 0,
\end{cases}
\]
where $b > 0$ is the collective benefit and $c > 0$ is the provision cost. Throughout this work we use $b = 2$ and $c = 1$, following the parameterisation of~\cite{bib1}. The joint strategy distribution is
\[
q_{\pi_1,\ldots,\pi_n}(x_1,\ldots,x_n)
=
\prod_{i=1}^{n}
\pi_i^{x_i}(1-\pi_i)^{1-x_i}.
\]
Players are assumed to choose independently, so the distribution factorizes. This formulation naturally interpolates between pure strategies: $\pi_i = 1$ corresponds to volunteering with certainty, while $\pi_i = 0$ represents full abstention.

Unlike the pure-strategy case, this game admits a unique symmetric Nash equilibrium. Specifically, there exists a unique value $\alpha_n \in (0,1)$ such that the profile $(\alpha_n, \ldots, \alpha_n)$ constitutes an equilibrium. This value is obtained as the unique root in $(0,1)$ of the degree-$n$ polynomial
\[
g_n(\alpha)
=
(1-\alpha)^{n-1}(2n\alpha + 1 - \alpha) - 1.
\]
This result, due to Weesie and Franzen~\cite{bib7}, implies that the probability of volunteering decreases with the number of players, with asymptotic behavior
\[
\alpha_n \sim \frac{\omega^*}{n} + \mathcal{O}(n^{-2}),
\]
where $\omega^*$ is the positive solution of $e^{\omega} = 1 + 2\omega$.

The expected payoff of each player at the symmetric equilibrium is:
\[
\$_i^{\mathrm{MVD}}(\alpha_n, \ldots, \alpha_n)
=
\left(2 - \frac{1}{n}\right)
\left[1 - (1 - \alpha_n)^n \right].
\]
This is the classical benchmark against which quantum payoffs are compared throughout the paper.

\subsection{The Quantum Volunteer's Dilemma}\label{subsec:quantum}

Following~\cite{bib1}, we extend the Volunteer's Dilemma to the quantum domain using the EWL quantization protocol~\cite{bib3}. In this framework, each player controls a qubit and may apply local unitary operations, expanding the strategy space beyond classical mixtures.

\bigskip
\noindent\textbf{Game setup}
\medskip

The game begins with a maximally entangled initial state of $n$ qubits:
\[
\ket{\psi_0} = J \ket{0}^{\otimes n},
\]
where the entangling operator $J$ is defined as
\[
J = e^{-i \frac{\pi}{4} Y^{\otimes n}} = \frac{1}{\sqrt{2}}(I - i Y^{\otimes n}),
\]
with $Y$ the Pauli-$Y$ matrix. Each player $i$ selects a quantum strategy represented by a unitary $U(\theta_i, \phi_i)$ from the family
\[
U(\theta, \phi) =
\begin{pmatrix}
e^{i\phi}\cos\!\left(\tfrac{\theta}{2}\right) & \sin\!\left(\tfrac{\theta}{2}\right) \\
-\sin\!\left(\tfrac{\theta}{2}\right) & e^{-i\phi}\cos\!\left(\tfrac{\theta}{2}\right)
\end{pmatrix},
\]
with $(\theta_i, \phi_i) \in [0,4\pi) \times [0,2\pi)$. The final state prior to measurement is
\[
\ket{\psi_f} = J^\dagger \left( \bigotimes_{i=1}^n U(\theta_i, \phi_i) \right) J \ket{0}^{\otimes n},
\]
where $J^\dagger = J^{-1}$ is the Hermitian adjoint of the entangling operator $J$, applied after the players' individual strategy choices. In our implementation, an additional $X$ gate is applied to each qubit before measurement to align the quantum encoding with the classical game convention, where $\ket{1}$ represents volunteering and $\ket{0}$ represents abstention.

\bigskip
\noindent\textbf{Outcome distribution and payoff}
\medskip

Measurement in the computational basis induces a probability distribution over classical strategy profiles $x \in \{0,1\}^n$:
\[
p_{\theta,\phi}(x) = |\langle x | \psi_f \rangle|^2.
\]
The expected payoff of each player is defined analogously to the classical case (Section~\ref{subsec:classical}):
\[
\$_i(\theta_1,\phi_1,\ldots,\theta_n,\phi_n)
=
\sum_{x \in \{0,1\}^n}
\$_i^{VD}(x)\, p_{\theta,\phi}(x).
\]

\bigskip
\noindent\textbf{Symmetric quantum Nash equilibrium}
\medskip

For $n \leq 9$, it has been shown in~\cite{bib1} that a symmetric Nash equilibrium exists in which all players adopt the same strategy $Q = (0, \pi/n)$, i.e., the profile $(Q, Q, \ldots, Q)$. This corresponds to the unitary
\[
U\!\left(0, \frac{\pi}{n}\right) =
\begin{pmatrix}
e^{\frac{i\pi}{n}} & 0 \\
0 & e^{-\frac{i\pi}{n}}
\end{pmatrix}.
\]
Under this profile, quantum interference induced by the entangling operator enhances the probability of configurations with multiple volunteers, in contrast to the classical equilibrium where the individual probability of contributing decreases with $n$. As a result, the expected payoff at this quantum equilibrium can exceed that of the classical symmetric equilibrium. How well this advantage holds up on real hardware is what the following sections assess.

\subsection{NISQ Hardware and Noise Sources}\label{subsec:nisq}

Current quantum processors operate in the \emph{Noisy Intermediate-Scale
Quantum} (NISQ) regime~\cite{bib5}, where circuit execution is subject to
multiple noise sources that degrade quantum correlations. These mechanisms
directly shift the probability distribution over strategy profiles and thus
determine whether game-theoretic predictions survive hardware execution.

\bigskip
\noindent\textbf{Gate errors and depth accumulation}
\medskip

Each quantum gate introduces a finite probability of error that accumulates
with circuit depth~\cite{bib13}. Two-qubit gates exhibit significantly
higher error rates than single-qubit gates and are the primary source
of degradation in circuits with entanglement~\cite{bib18}. In the EWL
protocol, the CNOT chain implementing $\exp\!\left(-i\frac{\pi}{4}
Z^{\otimes N}\right)$ grows linearly with $N$; this term is the
dominant contributor to hardware-induced fidelity loss.

\bigskip
\noindent\textbf{Decoherence: relaxation and dephasing}
\medskip

Decoherence describes the loss of quantum information due to interaction
with the environment, characterized by the relaxation time $T_1$ and
dephasing time $T_2$. For a gate of duration $t_g$, the system evolution
is modeled by a thermal relaxation channel parameterized by $(T_1, T_2,
t_g)$, introducing both population loss and coherence decay~\cite{bib13}.

\bigskip
\noindent\textbf{Crosstalk and residual couplings}
\medskip

Beyond the noise sources described above, NISQ devices exhibit
\emph{crosstalk}: operations on one qubit can induce parasitic rotations
on neighboring qubits through residual electromagnetic coupling~\cite{bib11,bib12}.
Unlike gate errors and decoherence, crosstalk is inherently non-local
and cannot be captured by independent per-qubit noise models. Its
magnitude and structure depend on the specific qubit layout and
calibration state of the device at the time of execution.

\bigskip
\noindent\textbf{Readout errors}
\medskip

The measurement process introduces classical misassignment errors, modeled
by confusion probabilities $P(0|1)$ and $P(1|0)$, which are directly
relevant here since the game payoff depends on the observed bitstring
distribution.

\bigskip
\noindent\textbf{The digital twin as a first-order approximation}
\medskip

To quantify the impact of these noise sources, we construct a
\textit{digital twin} of the device from real calibration data,
incorporating thermal relaxation, gate errors, and readout confusion
as independent per-qubit channels (Section~\ref{subsec:twin}). This
model is an effective first-order approximation: it captures the dominant, local noise mechanisms but does not reproduce non-local
effects such as crosstalk or temporal calibration drift. The comparison between digital twin predictions and experimental results therefore serves a dual purpose: validating global noise trends where the model is expected to hold, and revealing residual discrepancies that point toward the higher-order effects it omits. This distinction is central to the analysis in
Section~\ref{sec:results}.

\section{Methodology}\label{sec:methodology}

\subsection{Hardware and Execution Environment}\label{subsec:hardware}

All experiments were performed on the \texttt{ibm\_kingston} quantum backend, accessed via Qiskit Runtime with IBM Quantum API authentication~\cite{bib21}. This device belongs to the family of superconducting processors and employs ECR gates as the native two-qubit operation.

To reduce variability associated with hardware recalibrations, all circuit executions were carried out in a single experimental session using the same backend and calibration conditions. In the NISQ regime, parameters such as $T_1$, $T_2$ times and gate errors can fluctuate on timescales of the order of hours; session consistency is therefore a prerequisite for controlled comparison.

To implement the EWL protocol consistently across all system sizes ($N = 2, \ldots, 9$), we selected a subset of physically connected qubits forming a linear chain (qubits $0$ through $8$). This choice allows the interactions required by the entangling operator to be directly mapped without introducing additional SWAP gates, thereby reducing effective circuit depth and error accumulation.

Circuit execution was performed using Qiskit Runtime's \texttt{SamplerV2} primitive. Four independent jobs were submitted, one per transpiler optimization level, each containing circuits for all player counts ($N = 2, \ldots, 9$) repeated 20 times, grouped into Primitive Unified Blocs (PUBs) to improve communication efficiency and reduce execution latency. Each circuit was executed with 2048 shots per repetition, providing sufficient statistical resolution for the comparative analysis of fidelities and payoffs across the 20-sample distribution.

Post-processing readout error correction was applied using the \texttt{mthree} library~\cite{bib23,bib24}, calibrated from the confusion probabilities $P(0|1)$ and $P(1|0)$ extracted from the backend properties at the time of execution. This correction yields a more conservative estimate of target-state fidelity and is reported alongside raw results throughout Section~\ref{sec:results}.

\subsection{Circuit Construction}\label{subsec:circuit}

Using the Qiskit framework, we constructed the quantum circuit for the $N$-player game ($N = 2, \ldots, 9$) following the EWL quantization protocol of Koh--Kumar--Goh~\cite{bib1}. The circuit is structured into four main blocks:

\begin{enumerate}
    \item \textbf{Entangled state preparation:} the global operator $J$ is applied to $\ket{0}^{\otimes N}$. In this implementation it is decomposed as
    \[
    J = S^{\otimes N} H^{\otimes N} \exp\!\left(-i\frac{\pi}{4} Z^{\otimes N}\right) H^{\otimes N} S^{\dagger \otimes N},
    \]
    where $H$ is the Hadamard gate and $S$ is the phase gate. The non-local term $\exp\!\left(-i\frac{\pi}{4} Z^{\otimes N}\right)$ is implemented via a CNOT chain that propagates the global parity to a single qubit, followed by an $R_z(\pi/2)$ rotation and the reverse chain:
    \[
    \mathrm{CNOT}(0,1)\cdots \mathrm{CNOT}(N{-}2,N{-}1)\, R_z(\pi/2)\, \mathrm{CNOT}(N{-}2,N{-}1)\cdots \mathrm{CNOT}(0,1).
    \]
    This construction implements an effective $Z^{\otimes N}$ interaction using only single- and two-qubit gates.

    \item \textbf{Individual strategies:} each player locally applies the unitary
    \[
    U(\theta_i,\phi_i) = R_z(\pi - \phi_i)\, R_y(\theta_i)\, R_z(-\pi - \phi_i).
    \]
    All players adopt the symmetric Nash equilibrium strategy $(\theta_i,\phi_i) = (0,\pi/N)$, which reduces to a $Z$-axis phase rotation.

    \item \textbf{Inverse Operator $J^\dagger$:} applied by sequentially inverting the gates composing $J$.

    \item \textbf{Classical convention alignment:} an $X$ gate is applied to each qubit before measurement, so that $\ket{1}$ corresponds to volunteering.
\end{enumerate}

Under ideal conditions, this construction concentrates probability strongly on $\ket{1}^{\otimes N}$. However, the CNOT chain implementing $\exp\!\left(-i\frac{\pi}{4} Z^{\otimes N}\right)$ introduces a depth that grows with $N$; the resulting circuit is therefore sensitive to noise on NISQ devices.

\begin{figure}[htbp]
    \centering
    \includegraphics[width=\textwidth]{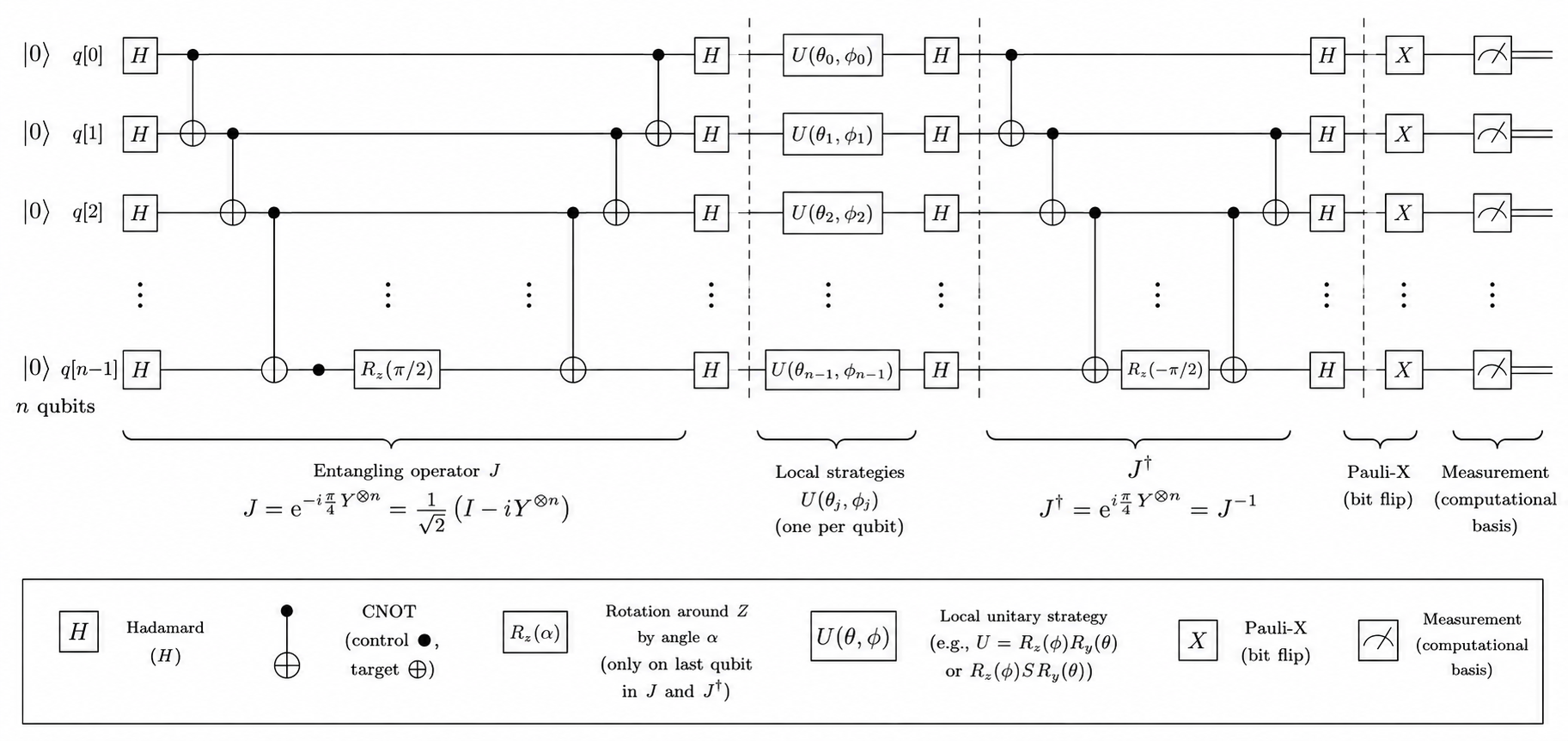}
    \caption{Quantum circuit implementing the $N$-player Quantum Volunteer's Dilemma following the EWL protocol. The circuit consists of: (i) an entangling operator $J = e^{-i\frac{\pi}{4}Y^{\otimes N}}$, decomposed into Hadamard, controlled-NOT, and $R_z$ rotations; (ii) local strategy unitaries $U(\theta_i,\phi_i)$ applied independently by each player; (iii) the inverse operator $J^\dagger$; and (iv) a final Pauli-$X$ layer to align measurement outcomes with the classical convention (volunteer = $\ket{1}$). Measurements are performed in the computational basis. The entangling layer and its inverse each contribute $2(N-1)$ CNOT/ECR gates, giving $4(N-1)$ two-qubit gates in total (ranging from 4 at $N=2$ to 32 at $N=9$). Qubits 0 through $N-1$ are mapped to a linear chain on \texttt{ibm\_kingston}.}
    \label{fig:Quantum_circuit_implementation}
\end{figure}

Figure~\ref{fig:Quantum_circuit_implementation} shows the circuit representation for the multiplayer case, illustrating the entanglement structure, local strategy application, and final alignment block.

\subsection{Transpiler Optimization Levels}\label{subsec:transpiler}

Each circuit was compiled using optimization levels 0 through 3 provided by Qiskit's transpiler via \texttt{generate\_preset\_pass\_manager}~\cite{bib22} (see Appendix~\ref{secA2} for a detailed description of each level). These levels correspond to progressively more aggressive transformation strategies:

\begin{itemize}
    \item \textbf{Level 0:} minimal compilation, preserving the original circuit structure.
    \item \textbf{Level 1:} basic optimizations including redundant gate cancellation and initial qubit mapping.
    \item \textbf{Level 2:} more sophisticated routing heuristics and gate-count reduction.
    \item \textbf{Level 3:} aggressive transformations including subcircuit resynthesis and advanced two-qubit gate reduction.
\end{itemize}

The comparison between optimization levels is relevant given the CNOT-chain structure of $J$, which makes depth, and consequently error accumulation, highly sensitive to compilation decisions. We do not assume higher levels yield uniformly better results; we evaluate their impact empirically on target-state fidelity and game payoff to identify trade-offs between circuit depth, structure, and error predictability. Results are discussed in Section~\ref{sec:results}.

\subsection{Digital Twin Construction}\label{subsec:twin}

To quantitatively assess noise impact, we constructed a \textit{digital twin} of the quantum device from real calibration data extracted from the backend used in the experiments.

\bigskip
\noindent\textbf{Calibration parameter extraction}
\medskip

For each physical qubit, we extracted the relaxation ($T_1$) and dephasing ($T_2$) times, single and two-qubit gate error rates, gate execution times ($t_g$), and readout confusion probabilities $P(0|1)$ and $P(1|0)$. All parameters were obtained from the backend properties of the same experimental job, so the noise model reflects the actual hardware conditions at the time of execution.

\bigskip
\noindent\textbf{Noise model construction}
\medskip

A composite noise model was defined using Qiskit's \texttt{AerSimulator}~\cite{bib19}, with three contributions:

\begin{itemize}
    \item \textbf{Single-qubit gates:} composition of a depolarizing channel (stochastic gate errors) and a thermal relaxation channel $(T_1, T_2, t_g)$.
    \item \textbf{Two-qubit gates (ECR):} for qubit pair $(i,j)$, individual thermal relaxation channels per qubit (parameterised by $T_1^{(i)}, T_2^{(i)}, t_g^{(ij)}$) composed with a correlated $ZZ$ Pauli error channel
    \[
    \mathcal{E}_{ZZ}^{(ij)}(\rho) \;=\; (1 - \epsilon_{\mathrm{ECR},ij})\,\rho \;+\; \epsilon_{\mathrm{ECR},ij}\,(ZZ)\,\rho\,(ZZ),
    \]
    where $\epsilon_{\mathrm{ECR},ij}$ is the ECR gate error rate for pair $(i,j)$ extracted directly from backend properties via \texttt{job.properties().gate\_error("ecr", (i, j))}.
    \item \textbf{Readout errors:} confusion matrices from backend-reported $P(0|1)$ and $P(1|0)$ values.
\end{itemize}

The same circuits, transpilation procedure, and qubit layouts used on hardware were applied to the simulation, so differences between experimental and simulated results reflect noise model limitations exclusively. The \textit{digital twin} provides a first-order reference for trend identification and dominant error source estimation; it does not aim to reproduce all device-specific details such as spatial noise correlations or calibration drift.

\subsection{Payoff Evaluation Method}\label{subsec:payoff}

Results are expressed as count distributions over the computational basis and processed by a dedicated evaluation engine. For an outcome $x \in \{0,1\}^N$, the per-player payoff $\$_i^{VD}(x)$ is the deterministic payoff defined in Section~\ref{subsec:classical} with $b = 2$ and $c = 1$. Let $w(x)$ denote the number of volunteering players (bits equal to~1); the rule reduces to: $w(x) = 0$: all receive $0$; $w(x) > 0$: volunteers receive $2 - 1/w(x)$, abstainers receive $2$.

The expected payoff for player $i$ is:
\[
\Pi_i = \sum_{x \in \{0,1\}^N} p(x)\, \$_i^{\mathrm{VD}}(x),
\]
where $p(x)$ is the empirical probability estimated from measurement counts. Qiskit returns results in little-endian format, so bit order is reversed when processing each outcome to align with the player-qubit mapping.

From the probability distributions, the following metrics are computed for each $(N, \text{level})$ configuration:

\begin{itemize}
    \item \textbf{Target-state fidelity:} $P(\ket{1}^{\otimes N})$.
    \item \textbf{Mean probability per non-target state:} average of $p(x)$ over the $2^N - 1$ non-target outcomes.
    \item \textbf{Target-state weighted payoff contribution:} $P(|1\rangle^{\otimes N}) \cdot (b - c/N)$, representing the contribution of the target state to the total expected payoff.
    \item \textbf{Global average payoff:} expected payoff $\Pi_i$ across the full outcome distribution.
\end{itemize}

Statistical uncertainty is quantified using 95\% confidence intervals based on the $t$-Student distribution across the 20 independent repetitions:
$$\mathrm{CI}_{95} = t_{0.975,\,19} \cdot \frac{\sigma}{\sqrt{20}},$$
where $\sigma$ is the sample standard deviation across repetitions. For the target-state weighted payoff contribution, the confidence interval is derived analytically as $\mathrm{CI}_{\mathrm{target}} = (2 - 1/N) \cdot \mathrm{CI}_{P(|1\rangle^{\otimes N})}$, since this metric is an exact linear rescaling of the target-state probability.

\section{Experimental Results}\label{sec:results}

We present experimental results for the Quantum Volunteer's Dilemma on real hardware, alongside comparison with the digital twin, for $N = 2, \ldots, 9$ players. Section~\ref{subsec:fidelity} characterizes fidelity decay and noise structure across all four optimization levels and presents the gate count analysis that motivates the selection of L2 as the reference configuration. Section~\ref{subsec:hamming} analyzes the Hamming distance structure of measurement outcomes. Section~\ref{subsec:digital} validates the Digital Twin against hardware results. Section~\ref{subsec:Expected_payoff} presents the expected payoff analysis.

\subsection{Target-State Fidelity and Scaling}\label{subsec:fidelity}

\begin{figure}[ht]
    \centering
    \includegraphics[width=0.75\textwidth]{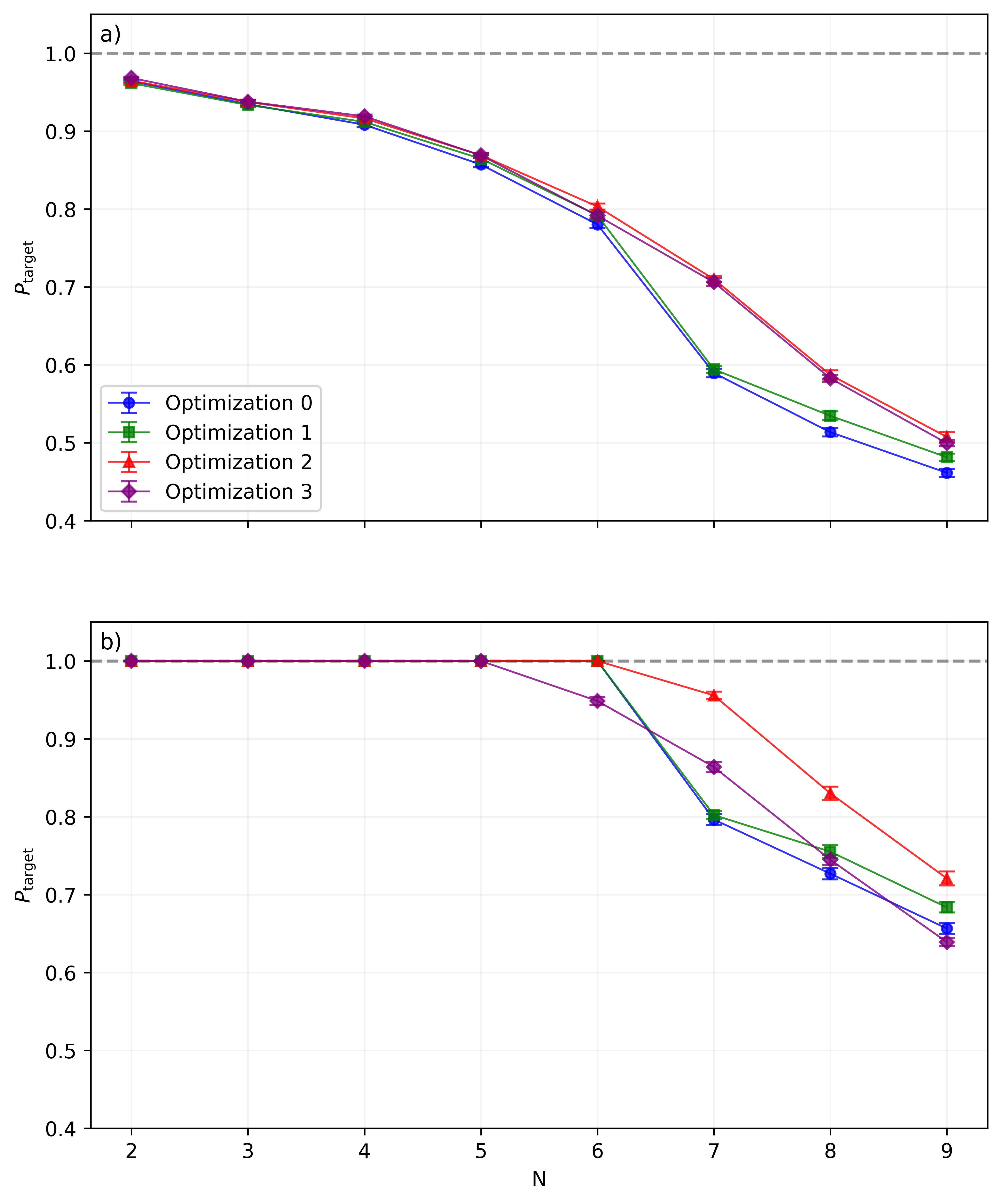}
    \caption{Target-state fidelity $P(|1\rangle^{\otimes N})$ as a function of the number of players for all four optimization levels. a) raw QPU measurements. b) readout-corrected measurements. Shaded bands represent 95\% $t$-Student confidence intervals across 20 repetitions.}\label{fig:target_state}
\end{figure}

Figure~\ref{fig:target_state} presents the target-state fidelity $P(|1\rangle^{\otimes N})$ as a function of player count for all four optimization levels, both without (Figure~\ref{fig:target_state}\,a) and with (Figure~\ref{fig:target_state}\,b) readout error correction. The decay is approximately quasi-linear in all cases (Figure~\ref{fig:target_state}\,a), with all levels producing comparable fidelities for $N \leq 5$.

Without readout correction, a divergence between levels becomes apparent from $N = 6$ onward: L0 and L1 exhibit a steeper and less regular decline, while L2 and L3 maintain a more stable and gradual decay through $N = 9$.

With readout correction applied, all four levels achieve near-complete target-state fidelity for $N \leq 5$. L3 begins to decay from $N = 6$, whereas L0, L1, and L2 remain stable until $N = 6$ before decreasing sharply. Among all levels, L2 exhibits the most stable trajectory and the least pronounced final decay, retaining the highest corrected fidelity at $N = 9$.

\textbf{Gate count justification.} Table~\ref{tab:gate_counts} reports the ECR gate count and circuit depth after transpilation for each level and player count. L2 and L3 each reduce the ECR gate count by 2 relative to L0 and L1 across all $N$, and produce identical compiled circuits for all player counts---confirming that the aggressive resynthesis of L3 yields no further structural improvement for this topology. The two additional gate reductions at L2, combined with its superior fidelity stability, justify its selection as the reference configuration.

\begin{table}[ht]
\caption{ECR gate count and circuit depth after transpilation for each optimization level (L0--L3) and player count ($N = 2, \ldots, 9$) on \texttt{ibm\_kingston}. Values obtained via \texttt{generate\_preset\_pass\_manager} with \texttt{initial\_layout} = $[0, 1, \ldots, N-1]$.}
\label{tab:gate_counts}
\centering
\begin{tabular}{c c c c c c c c c}
\toprule
\multirow{2}{*}{\textbf{N}} & \multicolumn{2}{c}{\textbf{L0}} & \multicolumn{2}{c}{\textbf{L1}} & \multicolumn{2}{c}{\textbf{L2}} & \multicolumn{2}{c}{\textbf{L3}} \\
\cmidrule(lr){2-3}
\cmidrule(lr){4-5} 
\cmidrule(lr){6-7} 
\cmidrule(lr){8-9}
& ECR & Depth & ECR & Depth & ECR & Depth & ECR & Depth \\
\midrule
2 & 4 & 55 & 4 & 28 & 2 & 14 & 2 & 14 \\
3 & 8 & 71 & 8 & 48 & 6 & 36 & 6 & 36 \\
4 & 12 & 87 & 12 & 62 & 10 & 46 & 10 & 46 \\
5 & 16 & 103 & 16 & 80 & 14 & 60 & 14 & 60 \\
6 & 20 & 119 & 20 & 96 & 18 & 72 & 18 & 72 \\
7 & 24 & 135 & 24 & 112 & 22 & 84 & 22 & 84 \\
8 & 28 & 151 & 28 & 128 & 26 & 96 & 26 & 96 \\
9 & 32 & 167 & 32 & 144 & 30 & 108 & 30 & 108 \\
\bottomrule
\end{tabular}
\end{table}

\subsection{Hamming Distance Analysis of Undesired States}\label{subsec:hamming}

\begin{figure}[ht]
    \centering
    \includegraphics[width=0.95\textwidth]{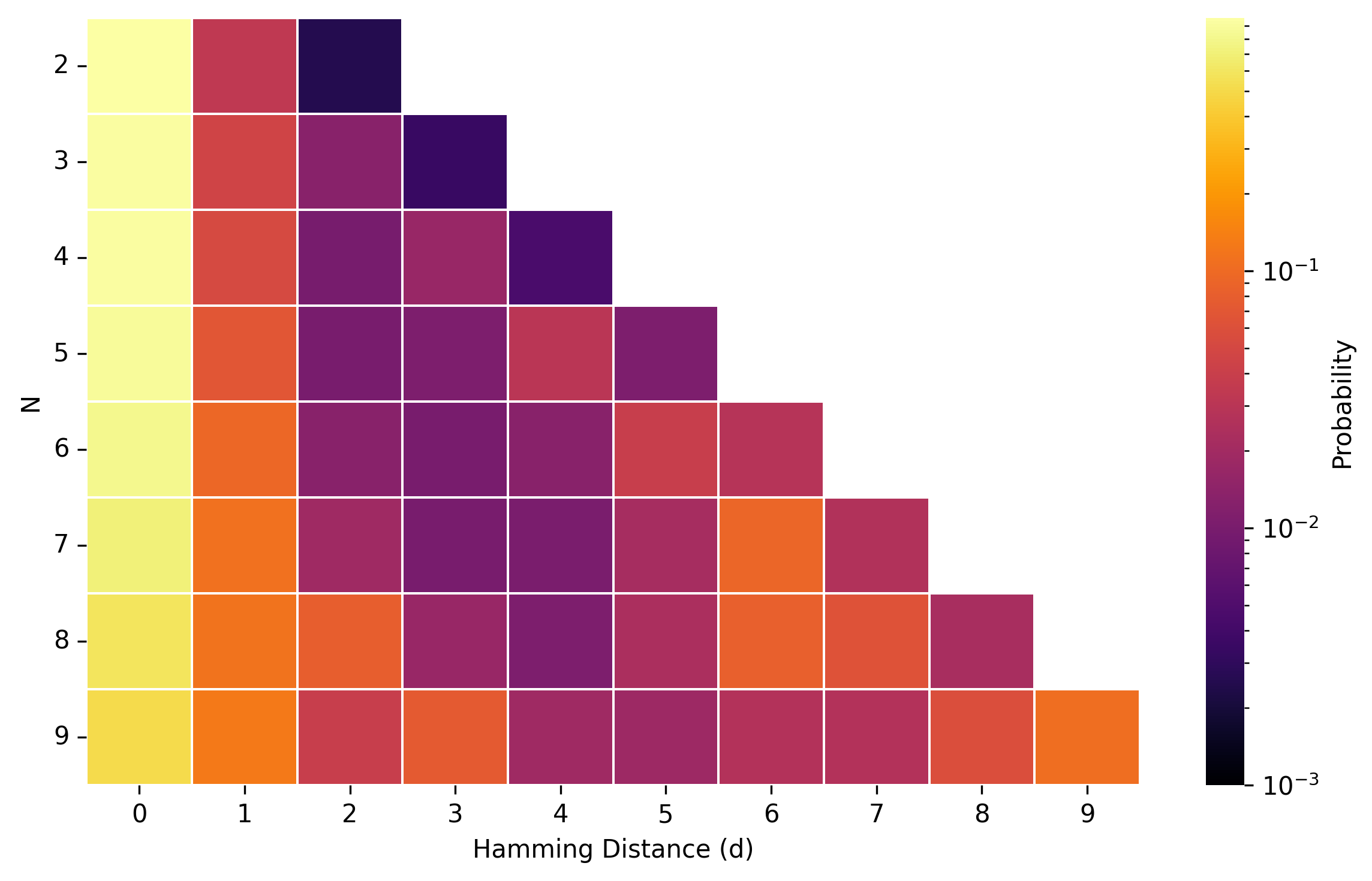}
    \caption{Heatmap of measurement outcome counts grouped by Hamming distance from the target state $|1\rangle^{\otimes N}$, aggregated over 20 repetitions at optimization level 2. Each row corresponds to a player count $N$; columns represent Hamming distances $d_H = 0, 1, \ldots, N$. Color intensity encodes normalized count frequency.}\label{fig:heatmap}
\end{figure}

Figure~\ref{fig:heatmap} presents a heatmap of the occurrence counts for all non-target measurement outcomes, grouped by Hamming distance $d_H$ from the target state $|1\rangle^{\otimes N}$. The analysis is performed on the raw (uncorrected) hardware measurement counts, aggregated over all 20 repetitions at optimization level L2. Readout-corrected and simulated counts are not used here: the Hamming structure reflects the physical error process as recorded by the device, before any post-processing redistribution of probabilities. The Hamming distance between two bitstrings $x, x_{\text{target}} \in \{0,1\}^N$ is defined as the number of positions at which they differ:
\begin{equation}
d_H(x, x_{\text{target}}) =
\sum_{i=1}^{N}
\mathbb{1}\left[x_i \neq x_{\text{target},i}\right],
\end{equation}
where $x_{\text{target}} = 11\ldots1$ denotes the theoretical equilibrium state.

In all player configurations, the distribution is dominated by $d_H = 0$ (the target state itself), with probability decreasing systematically for $d_H = 1, 2, \ldots$. This non-uniform structure, with states at distance 1 accumulating significantly more probability than those at larger distances, indicates that single-qubit errors dominate over correlated multi-qubit errors, providing empirical support for the independent per-qubit error channels used in the Digital Twin construction.

The heatmap further reveals that for $N \leq 6$, the target state retains a dominant probability and the redistribution toward $d_H \geq 1$ remains limited. From $N = 6$ onward, consistent with the fidelity behavior shown in Section~\ref{subsec:fidelity}, states at distances $d_H \geq 2$ begin to accumulate non-negligible counts, indicating a transition from single-qubit dominated errors to a regime where multi-qubit error contributions become relevant.

This redistribution explains why the target-state payoff contribution decays faster than the global average payoff: non-target states at $d_H = 1$ (a single abstainer among $N-1$ volunteers) still carry positive utility, partially compensating for the loss of $|1\rangle^{\otimes N}$ probability.

\subsection{Digital Twin Validation}\label{subsec:digital}

\begin{figure}[ht]
    \centering
    \includegraphics[width=0.95\textwidth]{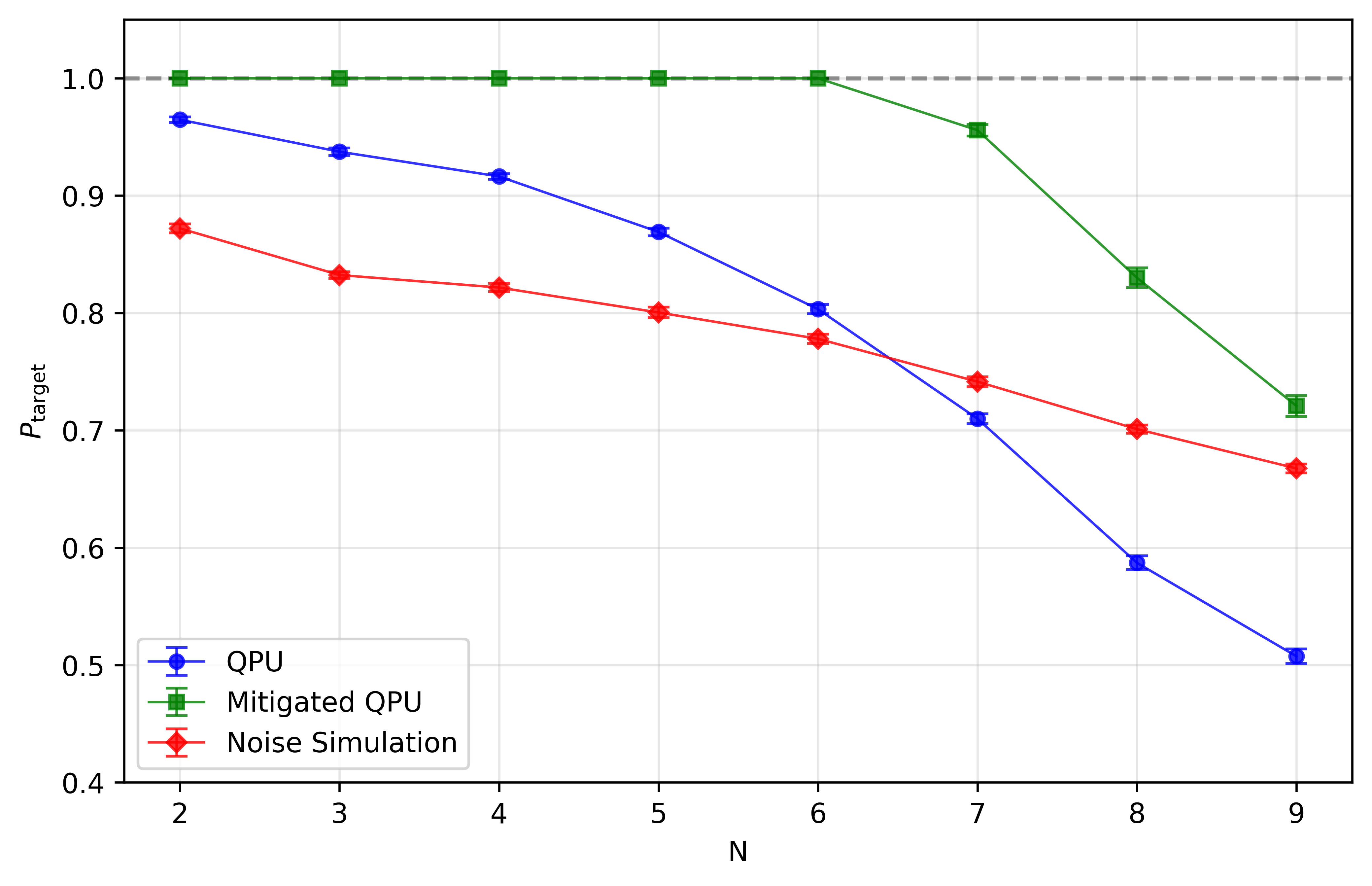}
    \caption{Target-state fidelity $P(|1\rangle^{\otimes N})$ as a function of player count for three regimes: raw QPU (L2), readout-corrected QPU (L2), and Digital Twin (mean over 20 seeds). Error bars represent 95\% $t$-Student confidence intervals.}\label{fig:comparison}
\end{figure}

Figure~\ref{fig:comparison} compares the target-state fidelity $P(|1\rangle^{\otimes N})$ obtained from raw QPU measurements, readout-corrected QPU measurements, and the Digital Twin simulation, all at optimization level L2. The Digital Twin was executed with 20 independent random seeds to assess model stability under stochastic variation; the low variance across seeds confirms that the model's stochastic component is small relative to its systematic noise structure.

For $N \leq 6$, the Digital Twin overestimates the noise present in the raw QPU data, that is, it predicts a lower fidelity than what is experimentally observed. For $N > 6$, the model transitions to underestimating the noise in the raw counts: the hardware degrades faster than the model predicts. 

The comparison between the Digital Twin and the readout-corrected QPU results deserves a methodological note: the Digital Twin incorporates readout confusion directly into the noise model via calibration-based confusion matrices, but no post-processing correction is applied to its output counts. The readout-corrected QPU results, by contrast, have undergone explicit probability redistribution via the \texttt{mthree} library. These two approaches address readout errors differently, so the comparison is not symmetric. Nevertheless, placing the Digital Twin alongside both raw and corrected QPU results is informative: it shows where the first-order noise model sits relative to the hardware before and after post-processing correction, and makes visible the fraction of the QPU-model gap that is attributable to readout noise specifically. With this framing, the observation that the Digital Twin consistently overestimates noise relative to the corrected data across the full range of $N$ indicates that the model's intrinsic error channels predict more degradation than what the hardware exhibits once readout misassignment is removed.

\subsection{Expected Payoffs}\label{subsec:Expected_payoff}

\begin{figure}[ht]
    \centering
    \includegraphics[width=0.75\textwidth]{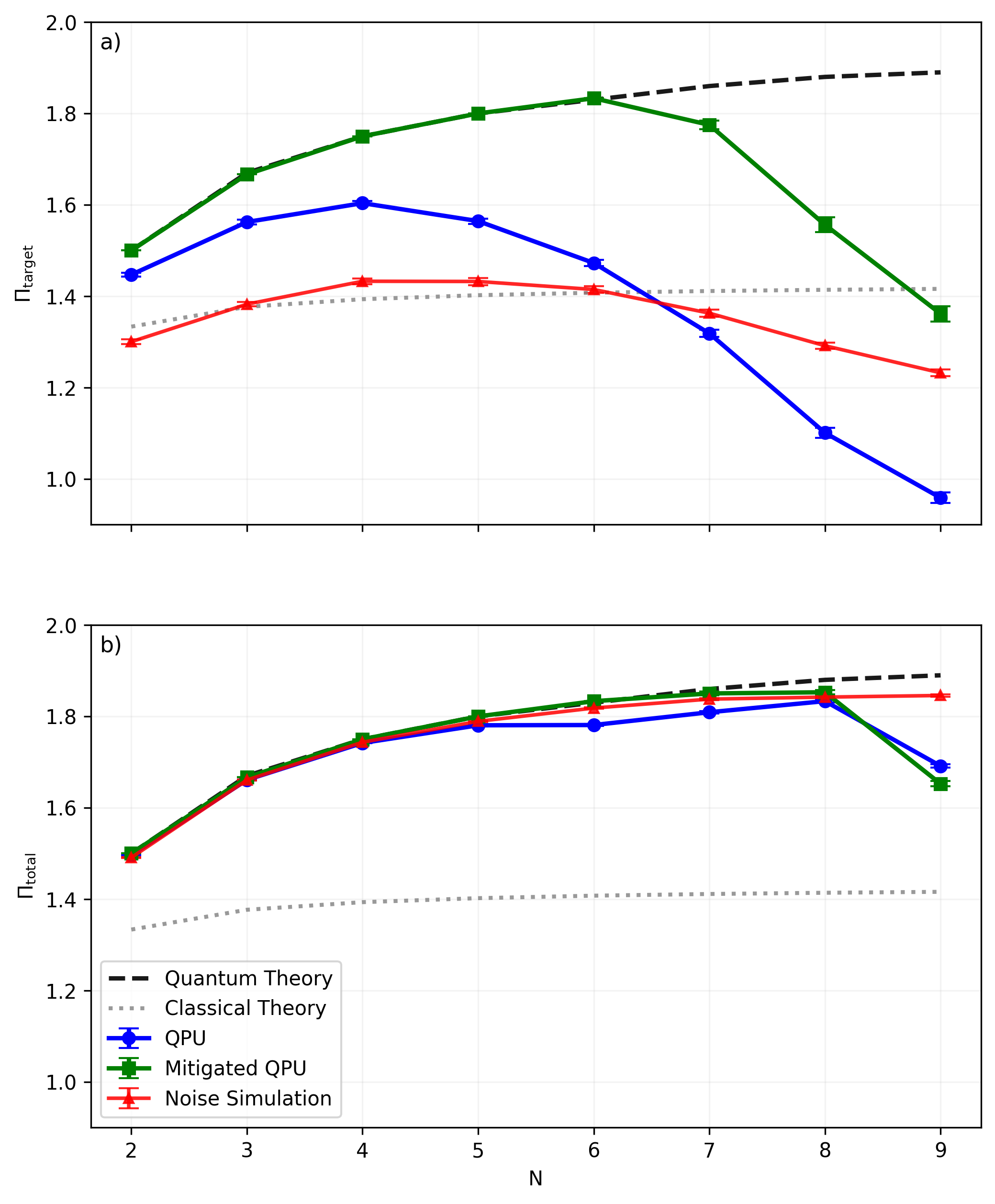}
    \caption{Expected payoffs at optimization level L2 as a function of player count. (a) Target-state weighted payoff contribution $P(|1\rangle^{\otimes N}) \cdot (b - c/N)$ for raw QPU, readout-corrected QPU, and Digital Twin, with theoretical benchmark and classical Nash equilibrium. (b) Global average payoff for the same four regimes. Error bars represent 95\% $t$-Student confidence intervals across 20 repetitions.}\label{fig:payment}
\end{figure}

Figure~\ref{fig:payment} presents two complementary but distinct payoff metrics at optimization level L2. Panel~(a) shows the target-state weighted payoff contribution $P(|1\rangle^{\otimes N}) \cdot (b - c/N)$: the share of the total expected payoff attributable solely to the $\ket{1}^{\otimes N}$ outcome, which decreases as that state loses probability to noise. Panel~(b) shows the global average payoff $\Pi_i$ summed over all $2^N$ outcomes: the full game-theoretic result that a player actually receives, which includes contributions from non-target states. These two metrics can behave very differently under noise: panel (a) tracks how well the quantum strategy concentrates probability on the ideal outcome, while panel (b) measures whether the overall cooperative structure of the game survives.
\\~\\
\noindent\textbf{Panel (a) -- Target-state contribution.} 
\\
Because this metric equals $P(\ket{1}^{\otimes N})$ scaled by a fixed factor $(b - c/N)$, it degrades in direct proportion to target-state fidelity. Both the raw and readout-corrected results exceed the classical Nash equilibrium up to $N = 6$. The corrected values approximately match the theoretical quantum benchmark for small $N$. Beyond $N = 6$, the raw payoff falls below the classical baseline, while the corrected values remain above it through $N = 8$, dropping below at $N = 9$. The Digital Twin remains below the classical equilibrium for most of the range, with brief exceptions at $N = 3$ and $N = 4$.
\\~\\
\noindent\textbf{Panel (b) -- Global average payoff.}
\\
\noindent All three regimes (raw, corrected, and Digital Twin) track the theoretical prediction closely up to $N = 5$. From $N = 5$ onward, the raw payoff decreases with its steepest drop at $N = 9$; the corrected values follow the theoretical curve through $N = 6$ before declining gradually. The Digital Twin decreases steadily from $N = 5$ without abrupt transitions. Across the full range, the global payoff of the raw and corrected hardware results remains above the classical Nash equilibrium, in contrast to the target-state contribution. This resilience arises from the positive utility of non-target states with at least one volunteer, as established by the Hamming distance structure in Section~\ref{subsec:hamming}.

Tables~\ref{tab:readout_impact} and~\ref{tab:error_metrics} report the per-player readout correction impact (absolute and relative offset between raw and corrected global payoffs) and the absolute and relative error metrics (AE and RE) for raw, readout-corrected, and Digital Twin results against the theoretical benchmark, respectively.

\begin{table}[ht]
\centering
\caption{Per-player readout correction impact: absolute and relative
offset between raw and readout-corrected global average payoff at
optimization level L2.}
\label{tab:readout_impact}
\begin{tabular}{ccc}
\toprule
N & Abs. offset & Rel. offset ($\%$) \\ \midrule
2 & 0.0038 & 0.25\\
3 & 0.0056 & 0.34\\
4 & 0.0081 & 0.47\\
5 & 0.0195 & 1.09\\
6 & 0.0522 & 2.93\\
7 & 0.0412 & 2.27\\
8 & 0.0195 & 1.06\\
9 & -0.0390 & -2.30\\ \bottomrule
\end{tabular}
\end{table}

\section{Discussion}\label{sec:discussion}

The central finding of this work is a systematic decoupling between state-level and aggregate-level behavior. With readout error correction, the global average payoff reproduces the quantum theoretical benchmark exactly for $N \leq 6$ and remains above the classical Nash equilibrium across the full range $N = 2, \ldots, 9$. This aggregate advantage persists even as target-state fidelity degrades substantially with system size, reaching approximately 51\% (raw) and 72\% (corrected) at $N = 9$. The payoff structure of the Volunteer's Dilemma, in which $2^N - 1$ of the $2^N$ possible outcomes contribute positively to expected utility, is what sustains this advantage under noise.

\subsection{Persistence of Quantum Advantage Under Noise}

Two distinct thresholds emerge from the data, depending on the metric considered. For the \emph{global average payoff}, readout-corrected results remain above the classical Nash equilibrium across the entire tested range ($N = 2, \ldots, 9$), with confidence intervals that do not overlap the classical baseline at any player count. Raw results also exceed the classical equilibrium for $N \leq 8$; at $N = 9$ the raw global payoff drops to $1.691 \pm 0.004$, which still clears the classical value of $1.416$ by a statistically significant margin.

For the \emph{target-state weighted payoff contribution}, the picture is more restrictive. Raw results fall below the classical baseline at $N = 7$; readout-corrected results maintain the advantage through $N = 8$ but drop below the classical value at $N = 9$. This metric therefore sets a more conservative practical threshold: quantum advantage is unambiguously resolvable up to $N = 6$ by either metric, with the global payoff extending the observable range further.

The Hamming distance analysis (Section~\ref{subsec:hamming}) provides a structural explanation for this difference: probability redistributed to $d_H = 1$ states still contributes positive utility, acting as a natural damping mechanism for the global metric. The practical limit of quantum advantage therefore depends both on the total probability lost from the target state and on the specific metric used to assess it.

\subsection{Digital Twin as a Noise Modeling Tool}

The Digital Twin, as a first-order approximation built from independent per-qubit calibration channels, provides a useful but limited representation of the hardware behavior. Its primary strength lies in reproducing the qualitative structure of aggregate metrics: the global average payoff profile closely tracks the experimental trend, and the model correctly identifies the direction and approximate magnitude of noise-induced degradation.

However, the Digital Twin's representation of target-state fidelity is less accurate. The model predicts a strictly linear decay, while the hardware exhibits a more gradual decline at lower player counts followed by a sharper degradation beyond $N = 6$. Quantitatively, the model overestimates noise for small $N$ (predicting lower fidelity than observed) and underestimates it for large $N$ in the raw data; for readout-corrected data, it consistently overestimates noise across the full range. This asymmetric behavior suggests that the non-local noise mechanisms omitted by the model, primarily crosstalk and temporal calibration drift, have a growing impact as circuit depth increases with $N$.

Despite these limitations, the low variance of the Digital Twin across 20 independent simulation seeds confirms that the model's stochastic component is well controlled, and that the discrepancies with hardware are systematic rather than random. The Digital Twin therefore remains a useful reference for trend identification, hypothesis testing, and estimating the contribution of individual noise channels, while its limitations motivate the development of more sophisticated models as discussed in Section~\ref{subsec:future}.

\subsection{Implications for Quantum Game Theory}

The resilience of the global average payoff observed in Section~\ref{subsec:Expected_payoff} has a structural explanation rooted in the payoff geometry of the game. Any bitstring with at least one~1 contributes positively to the expected utility, so $2^N - 1$ of the $2^N$ possible outcomes pay positively. Under any noise model that spreads probability away from $\ket{1}^{\otimes N}$, the leaked probability mostly lands in states that still contribute. This payoff geometry, not a deep property of quantum mechanics, is what keeps the aggregate payoff well above the classical baseline.

The experiments do confirm that this payoff geometry is accessible on current hardware. The QPU global payoff exceeds the classical Nash equilibrium for all tested configurations, retaining its aggregate advantage over classical mixed strategies even at noise levels that reduce individual-outcome fidelity to approximately 51\% at $N = 9$. Identifying which game structures will share this property on near-term devices is a useful open question for quantum game theory.

\section{Conclusion}\label{sec:conclusion}

We have experimentally implemented the Quantum Volunteer's Dilemma for $N = 2, \ldots, 9$ players on the \texttt{ibm\_kingston} quantum processor, providing a systematic characterization of this game under realistic NISQ noise conditions. Four transpiler optimization levels were evaluated across 20 independent repetitions with 2048 shots each, and post-processing readout error correction was applied throughout.

Two observability thresholds emerge depending on the metric. For the global average payoff, readout-corrected results exceed the classical Nash equilibrium across the entire tested range ($N = 2, \ldots, 9$), confirming that aggregate quantum advantage is robust to the noise levels of current NISQ hardware. For the target-state weighted payoff contribution, the advantage holds through $N = 8$ with correction and through $N = 6$ without. The $N = 6$ boundary thus marks the point beyond which single-qubit dominated errors transition to a regime where multi-qubit contributions become relevant, rather than the disappearance of aggregate advantage.

The Hamming distance analysis of raw measurement counts reveals a non-uniform error redistribution dominated by single-qubit flips, consistent with the independent per-qubit noise channels of the Digital Twin and explaining why the global payoff is more resilient than target-state fidelity. Gate count analysis confirms that optimization level 2 minimizes ECR gates while providing the most stable fidelity across the full player range; level 3 produces identical transpiled circuits and offers no additional benefit for this topology.

The Digital Twin captures global aggregate trends but systematically misrepresents target-state fidelity, particularly at large $N$, revealing the limitations of first-order noise models in capturing non-local and depth-dependent hardware effects.

\subsection{Future Work}\label{subsec:future}

The results obtained here open several directions for future research. The readout correction applied in this work represents only a minimal level of error mitigation; more advanced techniques~\cite{cai2023quantum}, including zero-noise extrapolation (ZNE) \cite{temme2017error,li2017efficient,bib15} and probabilistic error cancellation (PEC) \cite{temme2017error,endo2018practical}, could improve the accuracy of absolute payoff estimates and extend the range of player numbers over which quantum advantage is statistically resolvable. Extending the experiment to larger games in a practically meaningful way would also require controlling the noise accumulated by the deeper EWL circuits, either through lower-depth implementations of the entangling operator or through hardware with lower gate error rates.

The Digital Twin's limitations at large $N$ motivate the development of more sophisticated noise models that incorporate spatial qubit correlations, crosstalk between neighboring qubits in the CNOT chain, and temporal calibration drift. Such models would improve the predictive accuracy of the simulation and better characterize the transition from single-qubit dominated to multi-qubit correlated errors observed around $N = 6$.

Another direction is to extend the quantum framework to other variants of the Volunteer's Dilemma, including asymmetric models with player-dependent costs or benefits \cite{diekmann1993cooperation, weesie1993asymmetry, he2014evolutionary, healy2018cost, guo2023asymmetric}, threshold Volunteer's Dilemmas in which collective benefit requires sufficiently many volunteers \cite{chen2013shared,mago2023greed,riordan2026kthreshold}, timing variants with delayed volunteering \cite{weesie1993asymmetry,weesie1994incomplete}, repeated variants involving sustained interaction among players \cite{przepiorka2021emergence,kloosterman2023infinitely,amir2026repeated,mukhopadhyay2024repeated}, and cost-synergy models in which the volunteering cost decreases exponentially with the number of contributors \cite{amir2026volunteers}. Once such variants are formulated in the quantum setting, the experimental framework developed here could be used to test whether the payoff robustness observed in the Weesie--Franzen cost-sharing model~\cite{bib7} persists across different forms of the Volunteer's Dilemma. Adaptive versions of these games may also require mid-circuit measurements and feedforward \cite{hothem2025measuring}, requiring mitigation methods tailored to such circuits \cite{koh2026readout}.

Finally, applying this experimental framework to other multiplayer quantum games with different payoff structures \cite{khan2018quantum,toni2026compendium} would help identify which game-theoretic properties, beyond the specific cost-sharing mechanism studied here, favor noise robustness in NISQ implementations. This would help establish quantum games as a broader experimental setting for studying task-level robustness on NISQ hardware.

\backmatter

\bookmarksetup{startatroot}
\bmhead{Acknowledgements}

We acknowledge the use of IBM Quantum services for this work. The views expressed are those of the authors, and do not reflect the official policy or position of IBM or the IBM Quantum team.

\section*{Declarations}

\begin{itemize}
\item \textbf{Funding:} The authors received no external funding for this work.
\item \textbf{Conflict of interest:} The authors declare no competing interests.
\item \textbf{Data availability:} All experimental datasets, backend calibration files, and stored job results are publicly available at \url{https://github.com/GermanDarDiaz/Quantum-Volunteer-s-Dilemma}.
\item \textbf{Code availability:} The complete Qiskit implementation, including circuit construction, noise modeling, and statistical analysis, is publicly available at \url{https://github.com/GermanDarDiaz/Quantum-Volunteer-s-Dilemma}.
\item \textbf{Author contribution:} G.D.D.A.\ performed the experimental implementation and data analysis. J.A.A.H.\ performed formal analysis and manuscript review. C.A.D.\ coordinated the team and contributed to circuit design. S.A.C.O.\ contributed to the manuscript and research coordination.  N.D.H.\ contributed to quantum game analysis.  S.T.G.\ contributed to circuit design. D.E.K.\ provided the theoretical foundation.
\item \textbf{Related work:} This article is partially based on the undergraduate thesis work of G.D.D.A. (in preparation, Universidad del Cauca). The present manuscript extends that work with additional experimental results, larger system sizes, and collaborative contributions.
\end{itemize}

\begin{appendices}

\section{Complete Numerical Results}\label{secA1}


\begin{table*}[!ht]
\caption{Target-state probability results for optimization level L2. Reported values correspond to the mean over 20 executions with 95\% confidence intervals.}
\label{tab:l2_target_probability}
\centering
\scriptsize
\setlength{\tabcolsep}{6pt}
\begin{tabular}{c ccc}
\toprule
\multirow{2}{*}{$N$} &
\multicolumn{3}{c}{$P_{\mathrm{target}}$} \\
\cmidrule(lr){2-4}
&
Raw &
Mitigated &
Digital Twin \\
\midrule

2 &
$0.9647 \pm 0.0025$ &
$1.0000 \pm 1.78\times10^{-8}$ &
$0.8721 \pm 0.0037$
\\

3 &
$0.9374 \pm 0.0032$ &
$1.0000 \pm 2.27\times10^{-8}$ &
$0.8323 \pm 0.0029$
\\

4 &
$0.9164 \pm 0.0024$ &
$1.0000 \pm 3.04\times10^{-8}$ &
$0.8217 \pm 0.0036$
\\

5 &
$0.8691 \pm 0.0032$ &
$1.0000 \pm 3.78\times10^{-8}$ &
$0.8004 \pm 0.0045$
\\

6 &
$0.8033 \pm 0.0039$ &
$1.0000 \pm 4.23\times10^{-8}$ &
$0.7780 \pm 0.0040$
\\

7 &
$0.7099 \pm 0.0042$ &
$0.9558 \pm 0.0049$ &
$0.7415 \pm 0.0043$
\\

8 &
$0.5871 \pm 0.0059$ &
$0.8300 \pm 0.0085$ &
$0.7010 \pm 0.0034$
\\

9 &
$0.5075 \pm 0.0062$ &
$0.7208 \pm 0.0089$ &
$0.6676 \pm 0.0038$
\\

\bottomrule
\end{tabular}
\end{table*}


\begin{table*}[!ht]
\caption{Target-state weighted payoff contribution associated with the target state for optimization level L2. Reported values correspond to the mean over 20 executions with 95\% confidence intervals.}
\label{tab:l2_conditional_payoff}
\centering
\scriptsize
\setlength{\tabcolsep}{6pt}
\begin{tabular}{c cc ccc}
\toprule
\multirow{2}{*}{$N$} &
\multicolumn{2}{c}{Benchmark} &
\multicolumn{3}{c}{Conditional payoff} \\
\cmidrule(lr){2-3}
\cmidrule(lr){4-6}
&
$Q_{\mathrm{th}}$ &
$Q_{\mathrm{cl}}$ &
Raw &
Mitigated &
Digital Twin \\
\midrule

2 &
1.50 & 1.333 &
$1.4470 \pm 0.0037$ &
$1.5000 \pm 2.67\times10^{-8}$ &
$1.3003 \pm 0.0056$
\\

3 &
1.67 & 1.377 &
$1.5623 \pm 0.0053$ &
$1.6667 \pm 3.78\times10^{-8}$ &
$1.3824 \pm 0.0048$
\\

4 &
1.75 & 1.393 &
$1.6037 \pm 0.0042$ &
$1.7500 \pm 5.32\times10^{-8}$ &
$1.4326 \pm 0.0063$
\\

5 &
1.80 & 1.402 &
$1.5643 \pm 0.0057$ &
$1.8000 \pm 6.81\times10^{-8}$ &
$1.4322 \pm 0.0080$
\\

6 &
1.83 & 1.408 &
$1.4727 \pm 0.0072$ &
$1.8333 \pm 7.76\times10^{-8}$ &
$1.4146 \pm 0.0073$
\\

7 &
1.86 & 1.411 &
$1.3185 \pm 0.0078$ &
$1.7750 \pm 0.0092$ &
$1.3628 \pm 0.0079$
\\

8 &
1.88 & 1.414 &
$1.1009 \pm 0.0110$ &
$1.5563 \pm 0.0160$ &
$1.2915 \pm 0.0064$
\\

9 &
1.89 & 1.416 &
$0.9586 \pm 0.0117$ &
$1.3615 \pm 0.0168$ &
$1.2322 \pm 0.0072$
\\

\bottomrule
\end{tabular}
\end{table*}


\begin{table*}[!ht]
\caption{Global payoff results for optimization level L2 considering all measured basis states. Reported values correspond to the mean over 20 executions with 95\% confidence intervals.}
\label{tab:l2_global_payoff}
\centering
\scriptsize
\setlength{\tabcolsep}{5pt}
\begin{tabular}{c cc ccc}
\toprule
\multirow{2}{*}{$N$} &
\multicolumn{2}{c}{Benchmark} &
\multicolumn{3}{c}{Global payoff} \\
\cmidrule(lr){2-3}
\cmidrule(lr){4-6}
&
$Q_{\mathrm{th}}$ &
$Q_{\mathrm{cl}}$ &
Raw &
Mitigated &
Digital Twin \\
\midrule

2 &
1.50 & 1.333 &
$1.4962 \pm 0.0008$ &
$1.5000 \pm 2.67\times10^{-8}$ &
$1.4910 \pm 0.0011$
\\

3 &
1.67 & 1.377 &
$1.6611 \pm 0.0009$ &
$1.6667 \pm 3.78\times10^{-8}$ &
$1.6608 \pm 0.0011$
\\

4 &
1.75 & 1.393 &
$1.7419 \pm 0.0009$ &
$1.7500 \pm 5.32\times10^{-8}$ &
$1.7435 \pm 0.0011$
\\

5 &
1.80 & 1.402 &
$1.7805 \pm 0.0020$ &
$1.8000 \pm 6.81\times10^{-8}$ &
$1.7892 \pm 0.0016$
\\

6 &
1.83 & 1.408 &
$1.7811 \pm 0.0024$ &
$1.8333 \pm 7.76\times10^{-8}$ &
$1.8183 \pm 0.0014$
\\

7 &
1.86 & 1.411 &
$1.8091 \pm 0.0026$ &
$1.8503 \pm 0.0040$ &
$1.8379 \pm 0.0018$
\\

8 &
1.88 & 1.414 &
$1.8334 \pm 0.0033$ &
$1.8529 \pm 0.0046$ &
$1.8423 \pm 0.0020$
\\

9 &
1.89 & 1.416 &
$1.6914 \pm 0.0039$ &
$1.6525 \pm 0.0058$ &
$1.8458 \pm 0.0026$
\\

\bottomrule
\end{tabular}
\end{table*}


\begin{table*}[!ht]
\centering
\caption{Absolute error (AE) and relative error (RE) with respect to the theoretical quantum benchmark for the raw QPU results, readout-corrected results, and Digital Twin simulation. The final columns quantify the impact of readout correction relative to the raw experimental measurements.}
\label{tab:error_metrics}
\renewcommand{\arraystretch}{1.15}
\begin{tabular}{c c c c c c c}
\toprule
\multirow{2}{*}{$N$} &
\multicolumn{2}{c}{\textbf{Raw QPU}} &
\multicolumn{2}{c}{\textbf{Readout Corrected}} &
\multicolumn{2}{c}{\textbf{Digital Twin}} \\
\cmidrule(lr){2-3}
\cmidrule(lr){4-5}
\cmidrule(lr){6-7}
& AE & RE (\%) & AE & RE (\%) & AE & RE (\%) \\
\midrule
2 & 0.0038 & 0.2515 & 0.0000 & 0.0000 & 0.0090 & 0.6030  \\
3 & 0.0089 & 0.5358 & 0.0033 & 0.1996 & 0.0092 & 0.5505  \\
4 & 0.0081 & 0.4639 & 0.0000 & 0.0000 & 0.0065 & 0.3711  \\
5 & 0.0195 & 1.0815 & 0.0000 & 0.0000 & 0.0108 & 0.5981  \\
6 & 0.0489 & 2.6697 & 0.0033 & 0.1822 & 0.0117 & 0.6397  \\
7 & 0.0509 & 2.7351 & 0.0097 & 0.5223 & 0.0221 & 1.1896  \\
8 & 0.0466 & 2.4769 & 0.0271 & 1.4407 & 0.0377 & 2.0069  \\
9 & 0.1986 & 10.5067 & 0.2375 & 12.5686 & 0.0442 & 2.3377  \\
\bottomrule
\end{tabular}
\end{table*}

\clearpage
\section{Transpiler Optimization Levels}\label{secA2}

The Qiskit transpiler accepts an \texttt{optimization\_level} parameter (integer, 0 to 3) that controls the degree of circuit transformation applied before hardware execution~\cite{bib2}. Level~0 applies no significant structural optimizations and preserves the original circuit structure. Level~1 introduces basic optimizations including redundant gate cancellation and initial mapping improvements. Level~2 incorporates additional heuristics for qubit assignment and gate reduction to better adapt the circuit to the backend topology. Level~3 employs more aggressive transformations including resynthesis and advanced two-qubit gate reduction. Higher optimization levels generally reduce effective circuit complexity, potentially decreasing error accumulation and improving experimental performance.

\end{appendices}

\bibliography{sn-bibliography}

\end{document}